\pdfoutput=1

\def\40K{$^{40}$K }
\def\K{$^{39}$K }
\def\Na{$^{23}$Na }

\def\Xstate{X$^1\Sigma^+$}
\def\astate{a$^3\Sigma^+$}

\documentclass[aps,pra,twocolumn,floatfix,10pt]{revtex4-1}

\usepackage{graphicx,amsmath,amssymb,ulem}
\usepackage{hyperref}
\usepackage{color}
\usepackage{epstopdf}
\usepackage{nicefrac}
\usepackage[caption=false]{subfig}
\usepackage{gensymb}
\usepackage{amsmath}
\usepackage{siunitx}
\usepackage{soul}

\newcommand\Tstrut{\rule{0pt}{3.6ex}}        
\newcommand\Bstrut{\rule[-1.9ex]{0pt}{0pt}}   

\usepackage{xstring}

\usepackage{multirow}
\usepackage{array}
\usepackage{booktabs}

\newcommand{\executeiffilenewer}[3]{            
	\ifnum\pdfstrcmp{\pdffilemoddate{#1}}           
	{\pdffilemoddate{#2}}>0
	{\immediate\write18{#3 }}\fi            
}

\def\K{$^{39}$K }
\def\Na{$^{23}$Na }

\begin{document}
	
	\title{Feshbach resonances in $\boldsymbol{^{23}\mathrm{Na}+}$$\boldsymbol{^{39}\mathrm{K}}$ mixtures and refined molecular potentials for the NaK molecule}
	
	\author{Torsten~Hartmann}
	\author{Torben A.~Schulze}
	\author{Kai~K.~Voges}
	\author{Philipp~Gersema}
	\author{Matthias~W.~Gempel}
	\author{Eberhard~Tiemann}
	\author{Alessandro~Zenesini}
	\author{Silke~Ospelkaus}
	\email{silke.ospelkaus@iqo.uni-hannover.de}

	\affiliation{Institut f\"ur Quantenoptik, Leibniz Universit\"at Hannover, 30167~Hannover, Germany}
	
	\date{\today}
	
	\begin{abstract}
		
		We present a detailed study of interspecies Feshbach resonances of the bosonic \mbox{$^{23}\mathrm{Na}+$$^{39}\mathrm{K}$} mixture for magnetic fields up to $750 \, \mathrm{G}$ in various collision channels. A total of fourteen Feshbach resonances are reported, as well as four zero crossings of the scattering length and three inelastic two-body loss features. We use the observed magnetic field locations of the resonant features together with the known data on \mbox{$^{23}\mathrm{Na}+$$^{40}\mathrm{K}$} to refine the singlet and triplet ground state potentials of NaK and achieve a consistent description of Feshbach resonances for both, the Bose-Bose mixture of \mbox{$^{23}\mathrm{Na}+$$^{39}\mathrm{K}$} as well as the Bose-Fermi mixture of \mbox{$^{23}\mathrm{Na}+$$^{40}\mathrm{K}$}. We also discuss the influence of the interplay between inelastic two-body and three-body processes on the observation of a Feshbach resonance.
		
	\end{abstract}
	
	\maketitle
	\section{Introduction}
	\label{sec1}

	Mixtures of ultracold atoms have recently attracted great interest as they enable the study of exciting quantum many-body effects \cite{GrimmJulienneChengChinFBRReview}. Furthermore, molecules in their ro-vibrational ground state are promising candidates for the study of dipolar many-body physics \cite{ReviewDipolarPhysics2008,ReviewDipolarManyBody2009Lahaye,ReviewDipolarPhyiscs2012}. Ensembles of \mbox{$\mathrm{Na}+$$\mathrm{K}$} feature several predicted interspecies Feshbach resonances \cite{VielSimoni} and their heteronuclear molecules possess a large electric dipole moment of $2.72\,\text{Debye}$ \cite{DipoleMomentsDulieu2005}, the freedom to prepare both bosonic and fermionic NaK molecules and chemical stability in molecular two-body collisions \cite{GsDiMo23Na40K2015}. 
	
	Up to now, Feshbach resonances for the mixture of bosonic $^{23}\text{Na}$ and fermionic $^{40}\mathrm{K}$ have been reported \cite{FBR23Na40K20121,FBR23Na40K2017ChineseGroup1} and fermionic ground state \mbox{$^{23}\mathrm{Na}$$^{40}\mathrm{K}$} molecules have been prepared by association of Feshbach molecules \cite{FBMol23Na40K2012} and subsequently by following two different STIRAP paths \cite{GsDiMo23Na40K2015,GsDiMo23Na40KFerm2018Bloch}. Thanks to these results and to previously obtained spectroscopic data \cite{Gerdes2008,NaKSpectroscopyTiemann2015}, the interatomic molecular potentials of the singlet and triplet ground states have been refined \cite{FBR23Na40K2017ChineseGroup1}, leading to predictions for Feshbach resonances in the bosonic mixtures by isotope mass rescaling. However, the very different hyperfine coupling between $^{23}\text{Na}$ and $^{40}\mathrm{K}$ compared to $^{23}\text{Na}$ and $^{39}\mathrm{K}$ lead to different singlet-triplet mixing within the Feshbach manifold. The accuracy of the predictions strongly depend on residual correlations in the determination of the singlet and triplet potential from measured \mbox{$^{23}\mathrm{Na}+$$^{40}\mathrm{K}$} Feshbach resonances. A direct measurement of the Feshbach resonance positions in the bosonic mixture is necessary to further refine the potentials for NaK and in particular to minimize correlations between the singlet and triplet molecular potentials.

	Experimental investigations of the Feshbach resonance spectrum of the bosonic pair \mbox{$^{23}\mathrm{Na}+$$^{39}\mathrm{K}$}  have started recently. The identification of Feshbach resonances in the $\left|f=1,m_{f}=-1\right\rangle_{\mathrm{Na}} + \left|f=1,m_{f}=-1\right\rangle_{\mathrm{K}}$ channel has been the basis for the preparation of a dual-species Bose-Einstein condensate of $^{23}\text{Na}$ and $^{39}\mathrm{K}$ atoms in the vicinity of a Feshbach resonance at about $247 \, \mathrm{G}$ \cite{SchulzeBEC2018}. Here, $f$ denotes the total angular momentum of the respective atom and $m_{f}$ denotes the projection onto the quantization axis.
	By comparing the measured resonance positions with predictions by Viel \textit{et al.} \cite{VielSimoni} significant deviations became apparent, whereas predictions making use of the recent evaluation in \cite{FBR23Na40K2017ChineseGroup1}, including measured d-wave resonances of \mbox{$^{23}\mathrm{Na}+$$^{40}\mathrm{K}$}, reduce these deviations.
	Since the Feshbach resonances observed in the $\left|1,-1\right\rangle_{\mathrm{Na}} + \left|1,-1\right\rangle_{\mathrm{K}}$ mixture are only a small subset of the many possible resonances, the observed deviations have motivated a thorough search for the remaining structures in order to further improve the potential energy curves (PECs).
	
	In this paper we present a detailed study of Feshbach resonances in a variety of hyperfine combinations in the ground state manifold of \mbox{$^{23}\mathrm{Na}+$$^{39}\mathrm{K}$} for a magnetic field range from 0 to $750 \, \mathrm{G}$. Our approach follows the iterative procedure of prediction, measurement and model refinement that is typical for molecular spectroscopy. The paper is structured accordingly:
	
	In Sec.\,\ref{Over} we first give a brief summary of Feshbach resonance predictions for the bosonic \mbox{$^{23}\mathrm{Na}+$$^{39}\mathrm{K}$} mixture, which have been derived from molecular potentials obtained by conventional spectroscopy of the NaK molecule and measurements of Feshbach resonances in the Bose-Fermi mixture of \mbox{$^{23}\mathrm{Na}+$$^{40}\mathrm{K}$} by isotope mass rescaling.
	In the second part of the section the experimental sequence for the measurements is described.
	
	In Sec.\,\ref{Exper} we present our measurements of loss features arising from elastic and inelastic scattering resonances.
	
	In Sec.\,\ref{sec:Theory} we describe and discuss the updated molecular potentials and how they improve the current knowledge of the scattering properties of both the Bose-Bose and Bose-Fermi mixtures.
	
	In Sec.\,\ref{sec:InelasticLosses} a brief discussion of an inelastic loss feature observed in the  $\left| 1, 1 \right\rangle_{\mathrm{Na}}+\left| 1,  -1 \right\rangle_{\mathrm{K}}$ channel and its influence on the possible observation of a close lying Feshbach resonance is given. 
	
	\section{Theoretical and experimental overview}
	\label{Over}
	
	\subsection{Theoretical predictions from known molecular potentials and a brief discussion of atom-loss spectroscopy}
	\label{Theo}
	
	Molecular potentials for the \Xstate\, state and the \astate\, state, correlating with the atomic ground state asymptote, have been derived from extensive classical spectroscopy \cite{Gerdes2008,NaKSpectroscopyTiemann2015} and refined by measurements of Feshbach resonances of the atom pair \mbox{$^{23}\mathrm{Na}+$$^{40}\mathrm{K}$} \cite{FBR23Na40K20121,FBR23Na40K2017ChineseGroup1}. These can be used to predict Feshbach resonances for the pair \mbox{$^{23}\mathrm{Na}+$$^{39}\mathrm{K}$}, if the Born-Oppenheimer approximation is assumed to remain valid and by incorporating the proper hyperfine and Zeeman interaction of \K. Details about the potential representations and the interpretation of the coupled channel calculations are given in Sec.\,\ref{sec:Theory}.
	
	From the potentials we predict 32 resonances in the magnetic field region from 0 to $750 \, \mathrm{G}$ by coupled-channel calculations for atom pairs with magnetic quantum numbers $f_{\mathrm{Na}}=1$, $m_{f_\mathrm{Na}}=\{-1,0,+1\}$ and $f_{\mathrm{K}}=1$, $m_{f_\mathrm{K}}=\{-1,0,+1\}$. Additionally, resonances for $f=2$, $m_{f_\mathrm{Na,K}}=-2$ are predicted. The calculations are done for an ensemble temperature of $\SI{1}{\micro\kelvin}$ and therefore only s-wave resonances are considered. Comparing theoretical and experimental results in \cite{SchulzeBEC2018}, the accuracy of our predictions is expected to be on the order of a few Gauss, significantly simplifying the search in the experiment.

	We locate Feshbach resonance positions by atom-loss spectroscopy, making use of the strongly enhanced scattering length and associated large on-resonance three-body loss in the atomic ensemble \cite{efi}. The detected losses are referred to as an elastic loss signal throughout this paper. Atom loss is also observed in the case of inelastic two-body collisions, where the energy released by the spin exchange is sufficiently high to lead to a two-body loss process in the trapped sample. In both cases we identify the local maximum of the atom loss with a resonance position. Typically, three-body processes require higher particle densities than two-body processes to obtain comparable loss rates. Thus in our case inelastic loss processes are expected to elapse on a shorter time scale than the three-body loss and will therefore lead to strong losses already for short hold times. 
	
	Care has to be taken when investigating a zero crossing of the scattering length as function of magnetic field. The minimum of the detected losses is in general not identical to the zero of the scattering length \cite{Shotan2014}, and a measurement similar to Sec.\,\ref{expo} can give misleading results. 
	We instead localize this magnetic field position by exploiting the two-body losses that appear during optical evaporation, similar to the work in \cite{SchulzeBEC2018}. In this procedure the magnetic field strength is set to the target value before the start of the optical evaporation. Because the optical trapping potential depth ratio is $U_{\mathrm{K}}\approx2.51\, U_{\mathrm{Na}}$, predominantly $^{23}\mathrm{Na}$ is ejected from our crossed optical dipole trap (cODT), while $^{39}\mathrm{K}$ is sympathetically cooled. On a zero crossing two body collisions are suppressed and therefore also losses resulting from the evaporation process \cite{Amplitude}.

	\subsection{Experimental procedure}
	\label{expo}
	
	The experimental setup is described in detail in \cite{Gempel} and \cite{Schulze2018}. The experimental sequence is based on the experiences of \cite{SchulzeBEC2018}. Since a precise knowledge of the generated magnetic field strength is necessary, a calibration has been performed repetitively in the course of the measurement campaign. The calibration method is the same as already described in \cite{SchulzeBEC2018}. We use microwave spectroscopy on a sample of $^{23}\mathrm{Na}$ with a temperature of $\sim800\,\mathrm{nK}$ confined in the cODT. For a defined electric current we measure the microwave frequency of the $\left| f = 1, m_{f} = -1 \right\rangle\rightarrow\left| f = 2, m_{f} = 0 \right\rangle$ transition using the atom loss in the $\left| f = 1, m_{f} = -1 \right\rangle$ state as signal and calculate the corresponding magnetic field using the Breit-Rabi formula. The transition frequency is determined with an uncertainty of about $10\,\mathrm{kHz}$ leading to an uncertainty on the value for the magnetic field strength on the order of $30\,\mathrm{mG}$. This is typically small compared to the statistical uncertainty originating from the resonance loss measurements. In the following we give a brief summary of the experiment sequence and explain the applied modifications in comparison to \cite{SchulzeBEC2018}.
	
	First, an optically plugged magnetic quadrupole trap is loaded from a dual-species magneto-optical trap (MOT). The atoms are then transferred to a cODT where we prepare an ultracold mixture of $^{23}\text{Na}$ and $^{39}\mathrm{K}$, both in $\left| f = 1, m_{f} = -1 \right\rangle$, by optical evaporation. The temperature is $\sim\SI{1}{\micro\kelvin}$ for both species, as measured by time-of-flight (TOF) expansions.

	After the evaporation in the cODT has been completed, we transfer $^{23}\text{Na}$ and $^{39}\mathrm{K}$ to the spin-state combination of interest, making use of rapid-adiabatic-passage \cite{ARPdressedAtomInterpret1984} sequences. Their efficiency is close to unity and neither heating of the sample nor atom loss due to the transfers is observed in our experiment.
	
	For the atom loss spectroscopy, we vary the atom numbers of the two species, preparing one species as the majority component and the other one as the minority component. The peak densities in the cODT are between $1.5\cdot10^{12}\,\text{cm}^{-3}$ and $7.4\cdot10^{14}\,\text{cm}^{-3}$ for $^{23}\mathrm{Na}$ and between $3.9\cdot10^{13}\,\text{cm}^{-3}$ and $1.6\cdot10^{15}\,\text{cm}^{-3}$ for $^{39}\mathrm{K}$. Detected loss within the minority component provides the primary signal.
	We use different tuning knobs in the experimental sequence to adjust the atom numbers. The first one is given by the loading times of the dual-species MOT. The second tuning knob is the depth of the forced microwave evaporation we perform in our optically plugged magnetic quadrupole trap.
	Due to the smaller repulsion of $^{39}\mathrm{K}$ by the blue-detuned plug laser light, a deeper evaporation and thereby colder atomic sample leads to an increased $^{39}\mathrm{K}$ density close to the magnetic trap center compared to $^{23}\text{Na}$. This increases losses in the $^{39}\mathrm{K}$ cloud. Hence, the deeper the evaporation is performed the more the atom ratio inside the cODT is shifted towards a prevalence of $^{23}\mathrm{Na}$. 
	
	We ramp the magnetic field strength in a few milliseconds to the target value. The loss measurement is repeated multiple times for every magnetic field value. For every resonance under investigation we experimentally determine the appropriate holding time, ensuring that the minority cloud is not depleted completely at the minimum but the loss feature is well visible. The holding time varies between $10 \, \mathrm{ms}$ and $1000 \, \mathrm{ms}$ and its magnitude can be an indication whether inelastic two-body or inelastic three-body processes dominate the losses. 
	
	The number of remaining atoms is recorded by absorption imaging of the majority component in the cODT and of the minority component after a short TOF. Where possible, we ramp down the magnetic field to zero in $5$ to $40 \, \mathrm{ms}$ (depending on the initial magnetic field value) and image both species at zero magnetic field. 
	
	For some spin state combinations we find that low field Feshbach resonances and/or high background scattering lengths lead to sizable losses, rendering imaging both species at zero magnetic field unfavorable. To circumvent these additional losses, we perform high-field imaging on the $^{39}\mathrm{K}$ D$_2$-line. For magnetic field strengths beyond $200 \, \mathrm{G}$ the Paschen-Back regime is reached  for $^{39}\mathrm{K}$, where the electron angular momentum $(j, m_j)$ and the nuclear spin $(i, m_i)$ decouple from each other in both electronic states and instead align directly relative to the external magnetic field. With this, $f$ is no longer a good quantum number. Together with the selection rule $\Delta m_i = 0$, it is always possible to find a transition with $\Delta m_j = \pm1$, which then serves as closed imaging transition. We choose a magnetic field for which no other resonance has to be crossed and for which the scattering rate of the state combination under investigation is low. We then image $^{39}\mathrm{K}$ as the majority component in the cODT and follow the scheme above for the \Na detection. 
	
	To improve the signal-to-noise ratio of the atomic cloud pictures, especially for low atom numbers, the absorption images of $^{23}\text{Na}$ and $^{39}\mathrm{K}$ are post-processed. The background of every picture is reconstructed using an algorithm based on principal component analysis \cite{PCAonImagesOfColdAtomClouds2007} and this background is subtracted from the picture. The pictures taken at equal magnetic field values are then averaged and the resulting image is fitted with a two-dimensional Gaussian. From the fit, the atom number is derived. The errors on the atom number result from the standard deviations of the fit. They vary between the different resonance measurements because they incorporate shot-to-shot atom number fluctuations which can originate from the different required spin preparation steps and the number of resonances which need to be crossed to reach the magnetic field value under investigation. For lucidity they are only shown for two exemplary measurements in Fig.\,\ref{alltogether2}. The errors are propagated to the profile fit of the loss feature and therefore contribute to the uncertainties of the resonance positions (see Tab.\,\ref{tab:tableAllRes}).

	\section{Loss resonances and zero crossings}
	\label{Exper}

	\begin{figure*}[h]
		
		\centering
		\includegraphics*[width=.85\textwidth]{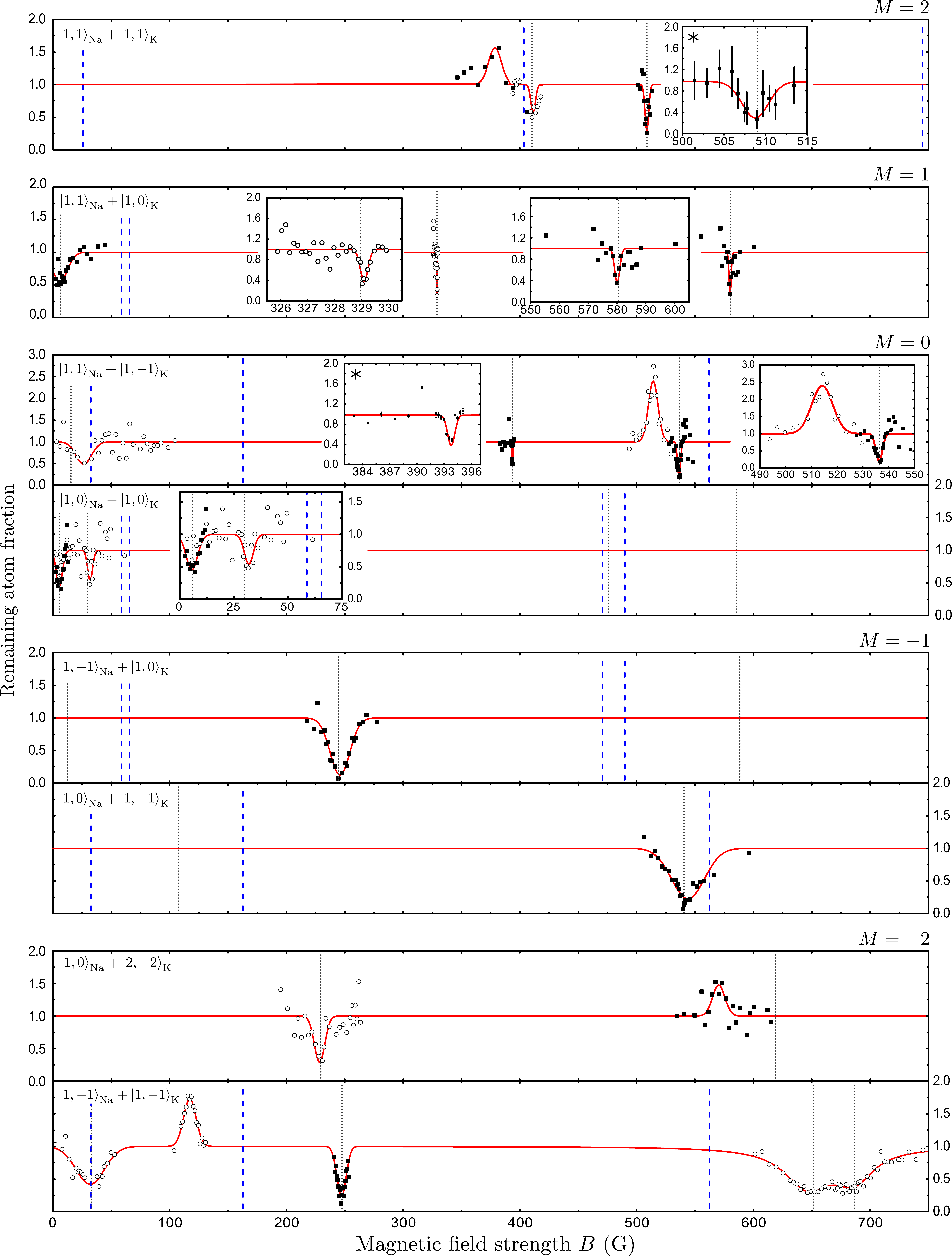}
		\caption{Collection of resonant features in different spin mixtures of \mbox{$^{23}\mathrm{Na}+$$^{39}\mathrm{K}$}. $M$ is the total magnetic quantum number of the pair $\left| f , m_{f} \right\rangle_{\text{Na}}$ + $\left| f , m_{f} \right\rangle_{\text{K}}$, $f$ is for increasing magnetic field strength $B$ only an approximate quantum number. Open circles and solid squares correspond to resonances observed by loss of $^{23}\text{Na}$ and of $^{39}\mathrm{K}$, respectively. Insets show zooms to the detected narrow resonance features. 	
			For two measurements, in $\left| 1 , 1 \right\rangle_{\text{Na}}$ + $\left| 1 , 1 \right\rangle_{\text{K}}$ and $\left| 1 , 1 \right\rangle_{\text{Na}}$ + $\left| 1 , -1 \right\rangle_{\text{K}}$ marked with (*), error bars are given, representing the variation in the errors for different loss measurements (for details, see text).
			For each recording, the holding time and the initial atom numbers are independently optimized. The data are normalized according to the respective phenomenological Gaussian fit (red solid line) of the feature with a baseline set to one. Zero crossings appear therefore artificially as enhancement of the atom number. Vertical gray dotted lines indicate the calculated positions of \mbox{$^{23}\mathrm{Na}+$$^{39}\mathrm{K}$} Feshbach resonances as listed in Tab.\,\ref{tab:tableAllRes}. Vertical dashed blue lines mark the positions of $^{39}\mathrm{K}$ resonances, taken from \cite{K39feshbach,FBR39K39KRoy2013}. The trace for the $\left|1,-1\right\rangle_{\mathrm{Na}} + \left|1,-1\right\rangle_{\mathrm{K}}$ mixture corresponds to data from Ref.\,\cite{SchulzeBEC2018}.
		}
		\label{alltogether2}
	\end{figure*}
	
	We have located 21 features, including the ones already presented in \cite{SchulzeBEC2018}. Fourteen features are assigned to predicted Feshbach resonances, resulting in three-body loss, three result from inelastic loss channels and four are assigned to zero crossings of the two-body scattering length. Figure\,\ref{alltogether2} shows the measured features. To determine the resonance positions we do a weighted fit to our data with a phenomenological Gaussian function. For every measurement the atom number is normalized on the baseline of the fit. In Fig.\,\ref{alltogether2}, the normalized atom numbers of the different measurements for every channel are set equal to one. This leads to an artificial increase in the atom number above one in case of a measured zero crossing. The resulting positions of all measured features are summarized in Tab.\,\ref{tab:tableAllRes}. In cases of loss features being clearly visible for both species, the position of the resonance is the weighted average of the center positions from the two fits. The error estimate of the experimentally determined resonance positions, given in Tab.\,\ref{tab:tableAllRes}, includes the uncertainty in the profile fit (which includes the errors from the atom number determination as explained above) as well as the uncertainty in the calibration of the magnetic field strength.
	
	As summarized in Tab.\,\ref{tab:tableAllRes}, some calculated resonances remain undetected. The main reasons are:
	\begin{itemize}
		\item Some of the state combinations experience a very high background scattering rate over the complete investigated magnetic field range. For these spin mixtures, the resonances are hidden since the atomic samples experience large losses already during the state preparation and/or the ramp to the target magnetic field.
		
		\item While pure $^{23}\text{Na}$ does not show significant loss features in the investigated range of the magnetic field strength, $^{39}\mathrm{K}$ exhibits several Feshbach resonances in different spin channels. We remeasured the $^{39}\mathrm{K}$ resonances relevant for our investigations and found all resonance positions to be within the experimental uncertainties of previous publications \cite{K39feshbach,FBR39K39KRoy2013}. Some of them are critical for our heteronuclear measurements since they are located near or even overlap with the resonance positions predicted for \mbox{$^{23}\mathrm{Na}+$$^{39}\mathrm{K}$}. These cases are mentioned in Tab.\,\ref{tab:tableAllRes}. Additionally, the $^{39}\mathrm{K}$ Feshbach resonances are indicated in Fig.\,\ref{alltogether2} as vertical blue dashed lines.
	\end{itemize}

	\begin{table*}[h]
		\caption{Measured magnetic field positions $B_{\text{exp}}$ and uncertainties $(\pm)$ together with calculated positions $B_{\text{th}}$, applying the improved potentials. $M$ is the total magnetic quantum number of the pair $\left| f , m_{f} \right\rangle_{\text{Na}}$ + $\left| f , m_{f} \right\rangle_{\text{K}}$, $f$ is in most cases only an approximate quantum number. Subscripts "$_{\text{res}}$" and "$_{\text{ZC}}$" stand for resonance and zero crossing, respectively. In some cases, maxima of the elastic (el.) and inelastic (in.) scattering rate are listed respectively. A.1 marks the inelastic loss feature discussed in section \ref{sec:InelasticLosses}. The measurements with (*) have been previously presented in \cite{SchulzeBEC2018}. }
		\label{tab:tableAllRes}	
		{\def\arraystretch{1.2}\tabcolsep=5pt
			\begin{tabular}{ c  c  c  c  c  c  c }
				\hline
				\hline
				$M$ & Na$_{f,m_{f}}$ & K$_{f,m_{f}}$ & $B_{\text{exp, ZC}}$ (G)  & $B_{\text{th, ZC}}$ (G)& $B_{\text{exp, res}} $ (G) &$B_{\text{th, res}}$ (G)\Tstrut\Bstrut \\
				\hline
				
				2 & 1,1& 1,1 & 380.88 (3.83) & 381.43 &411.33 (1.28) & 410.1\Tstrut\\
				& &  & - & 507.0 & 508.73 (0.83) & 508.81\Bstrut \\ 
				\hline 
				1 & 1,1& 1,0&  &  & 6.72 (2.09) &6.6\Tstrut\\ 
				& & &- & 328.5 &329.12 (0.77)  &  328.96 \\ 
				& & & -& 442.5 & close to KK res. & 467 \\
				& & &- & 577.5 & 579.94 (0.88) &  580.49\Bstrut \\ 
				& 1,0& 1,1&   &  &-  &7.5 (in.)  9.0 (el.) \\
				& & &- & 336.0 & close to KK res. &  419.0   \\ 
				& & &    &  & -& 508.5 (in.) 512.0 (el.)\Bstrut\\
				\hline

				0& 1,1& 1,-1&   &  & 26.34 (3.31) A.1 & 15.4 (el.) 28.2 (in.)\Tstrut\\
				& & & - & 393.0  & 393.61 (0.76) & 393.59  \\ 
				& & & 515.85 (1.68) & 516.4 & 536.07 (0.94) & 536.67 \Bstrut\\

				& 1,0& 1,0& - & 4.25 & 5.47 (1.01) & 5.6  \\ 
				& & &   &  & 31.86 (1.69)  & 29.8 \\
				& & & - &  407  & close to KK res. & 475.5 (in.) 476.0 (el.)\\
				& & &    &  &  -& 581.5 (in.) 585.5 (el.)\Bstrut \\

				& 1,-1& 1,1&   &  &- & 516.0 (el.) 522.5 (in.)\Bstrut\\

				\hline 
				-1 & 1,-1& 1,0&  &  & - &13.0\Tstrut\\  
				& & &   &  & 245.76 (1.45) & 244.75\\
				& & &  &  &- & 588.5 (el.) 593.0 (in.)\Bstrut \\
				& 1,0& 1,-1&   &  &- &107.5 \\ 
				& & &   &  & 541.09 (1.50)& 540.5\Bstrut \\
				& 1,1& 2,-2&   &  &-&  88.5 (in.)\\ 
				& & &   &  &- & 134.0 (in.) 138.0 (el.)\\ 
				& & &  &  & - & 471\Bstrut\\
				& 2,-2& 1,1&   &  & -&272.5 (in., weak) \\ 
				& & &  &  &-& 314.5 (in.)\\ 
				& & &  &  &- &465.5 (in.) 473.5 (el.) \Bstrut\\ 
				\hline 
				-2 & 1,-1& 1,-1& &  & 32.5 (0.8)(*) &33.13 \Tstrut\\
				& & &117.2 (0.2)(*) & 117.08  & 247.1 (0.2)(*) & 247.57 \\
				& & &  &  & 646.6 (1.5)(*)& 651.5 (el./in.)\\
				& & &  &  & 686.2 (1.5)(*)& 686.7 (in.)\Bstrut\\
				& 1,0& 2,-2&   &  &228.48 (1.49) & 229.5\\ 
				& & & 570.29 (2.55)& 574.3 &  -  &  619.0\Bstrut\\ 
				& 2,-2& 1,0&   &  &- & 358.5 (in.)\\   
				& & &  &  & - & 528.0 (in.) 533.0 (el.)\Bstrut\\  
				\hline
				\hline
			\end{tabular}
		}
	\end{table*}

	\section{Theory and Calculations}
	\label{sec:Theory}
	
	The theoretical modeling of two-body collisions of two alkali atoms in their electronic ground state is well established and described in many publications (see for example \cite{FBRofNaNa11Messurement2011}). The Hamiltonian contains the conventional kinetic and potential energy for the relative motion of the two particles and needs for the coupling of the molecular states \Xstate\, and \astate\, the hyperfine and Zeeman terms. For finer details of the partial waves with $l\geq 1$, one also needs the spin-spin interaction. The molecular PECs are represented in a power expansion of an appropriate function $\xi(R)$ of the internuclear separation $R$
	\begin{equation}
	\label{eq:xv} \xi(R)=\frac{R - R_m}{R + b\,R_m},
	\end{equation}
	to describe the anharmonic form of the potential function for $R\rightarrow \infty$ or $R\rightarrow0$. $R_m$ is an internuclear separation close to the minimum of the respective PEC and $b$ is a parameter to optimize the potential slopes left and right of $R_m$ with few terms in the power expansion. The full PECs are extended by long-range terms and short-range repulsive ones, see \cite{FBRofNaNa11Messurement2011} and the supplement of this paper.
	
	We calculate the two-body collision rate at the kinetic energy that corresponds to the temperature of the prepared ensemble. Thermal averaging is not performed, which would need significant computing time, but, more importantly, for a complete description we would have to consider the two cases of two- and three-body effects in the modeling. Here, we take the calculated maximum in the two-body collision rate constant to be equal to the observed Feshbach resonance and the calculated minimum to the observed loss minimum in optical evaporation.  
	
	The most recent fit of Feshbach resonances was reported for \mbox{$^{23}\mathrm{Na}+$$^{40}\mathrm{K}$} in \cite{FBR23Na40K2017ChineseGroup1} and the present evaluation starts from that result. Calculating the resonances for the observed cases with those derived PECs, we find significant deviations between observation and theory, thus demanding for new fits. They include all former observations and additionally the measured Feshbach resonances and zero crossings presented in this paper, in total 82 independent data points from Feshbach spectroscopy. This allows to further reduce correlations in the determination of the triplet and singlet potentials. The improved molecular potentials lead to a higher consistency between the measured and theoretically predicted resonance features for the bosonic \mbox{$^{23}\mathrm{Na}+$$^{39}\mathrm{K}$} mixture as well as the Bose-Fermi mixture of \mbox{$^{23}\mathrm{Na}+$$^{40}\mathrm{K}$}. The sum of squared residuals of calculated and experimentally determined resonance positions, weighted by the experimental uncertainties, improved from 337.0 to 255.31. We give a full listing of the data points and the evaluation with the different potential approaches in the supplement. Furthermore, we find that no inclusion of Born-Oppenheimer correction is needed to achieve this improvement. The parameters of the refined PECs can be found in the supplement.
	
	Table\,\ref{tab:tableAllRes} lists the calculated resonance positions which are derived using the updated potential energy curves.
	In several channels, the calculations show maxima for the elastic and inelastic scattering rates to appear close to each other. Such a constellation can lead to a shifted minimum in the atom-loss measurement. For one resonance a remarkable shift was observed in our experiment and will be discussed in the following section. Note that closely located maxima of elastic and inelastic scattering rates can also lead to asymmetric broadening of loss signals. This can be an additional reason why for several measurements of Feshbach resonances shifts and asymmetric broadening of the loss signals were reported \cite{FBRLineShapes2002,FBRLineShape2003,FBMolLineShape2008,EfimoveLineShape2012,AtomLossLineShape2011,FBRLineShape2012}.  
	
	\begin{figure}[h]
		\centering
		\includegraphics*[width=0.9\columnwidth]{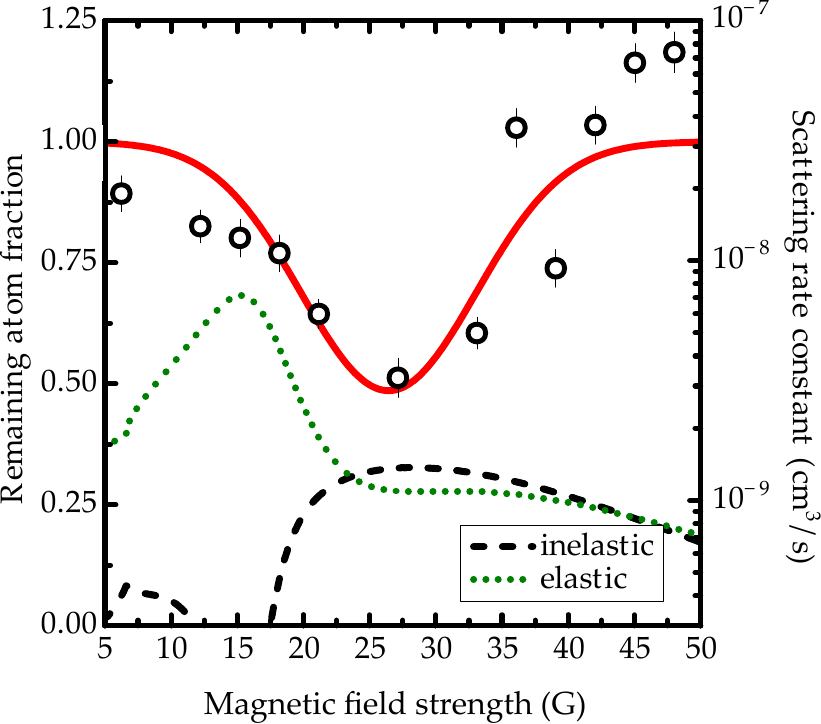}
		\caption{Remaining atom fraction of $^{23}\mathrm{Na}$ (open circles) and Gaussian fit (red curve) as well as elastic (green dotted line) and inelastic (black dashed line) collision rate constants for the $\left| 1, 1 \right\rangle_{\mathrm{Na}}+\left| 1,  -1 \right\rangle_{\mathrm{K}}$ channel. The rate constants are calculated for a kinetic energy of $\SI{1}{\micro\kelvin}$. The peak densities in the cODT for this measurement were $2.2\cdot10^{12}\,\text{cm}^{-3}$ for $^{23}\mathrm{Na}$ and $4.4\cdot10^{13}\,\text{cm}^{-3}$ for $^{39}\mathrm{K}$. }
		\label{figinel_1}
	\end{figure}
	
	\section{Inelastic loss feature in $\boldsymbol{\left| 1, 1 \right\rangle_{\mathrm{Na}}+\left| 1,  -1 \right\rangle_{\mathrm{K}}}$}
	\label{sec:InelasticLosses}

	Our theoretical model predicts a Feshbach resonance at $15\,\text{G}$ for the $\left| 1, 1 \right\rangle_{\mathrm{Na}}+\left| 1,  -1 \right\rangle_{\mathrm{K}}$ channel. However, our measurements show a broad loss signal at $26.34\,\text{G}$, see Fig.\,\ref{alltogether2} and Tab.\,\ref{tab:tableAllRes}. The large deviation can be explained by looking at both elastic and inelastic loss contributions. Figure\,\ref{figinel_1} shows the atom-loss measurement together with the elastic loss rate which has a maximum at 15.4\,G and an enhancement of the inelastic loss rate at $28.2\,\text{G}$. The inelastic part is due to the coupling to the $\left| 1, 0 \right\rangle_{\mathrm{Na}}+\left| 1,  0 \right\rangle_{\mathrm{K}}$ channel and gains strength through a Feshbach resonance at $29.8$\,G for the latter channel.
	The inelastic two-body losses occur on a shorter timescale and dominate the three-body losses invoked by the elastic part of the scattering rate.
	This is confirmed by a relatively short hold time of 100\,ms for the measurement presented in Fig.\,\ref{figinel_1}, in comparison to up to an order of magnitude larger hold times for the measurements of three-body losses at a Feshbach resonance.
	The elastic peak might be hidden in the shoulder of the recorded profile. To pinpoint the Feshbach resonance position with higher accuracy, a binding energy measurement should be performed \cite{LiCsEfimovTiemann2015} which is left for future investigation. Similar findings on the interplay of inelastic two-body and three-body processes have been reported in \cite{FBR23Na40K2017ChineseGroup1}.

	\section{Conclusion and Outlook}
	\label{sec:Result}
	
	In this paper we presented a detailed study of Feshbach resonances in many possible hyperfine combinations of the ground state manifold of \mbox{$^{23}\mathrm{Na}+$$^{39}\mathrm{K}$} in a magnetic field range from 0 to $750 \, \mathrm{G}$. We compared these measurements to theoretical predictions based on the currently available data for NaK molecular potentials and used our data to refine those potentials. The improved potentials lead to a higher consistency between the experimental data and theoretical predicted resonance features for both, the bosonic \mbox{$^{23}\mathrm{Na}+$$^{39}\mathrm{K}$} mixture and the Bose-Fermi mixture of \mbox{$^{23}\mathrm{Na}+$$^{40}\mathrm{K}$}. With the incorporation of the experimental data of the bosonic Feshbach resonances and the finding that the Born-Oppenheimer approximation remains valid for the investigated partial waves, reliable predictions based on the new potentials will be also possible for the other, still unexplored, bosonic mixture of \mbox{$^{23}\mathrm{Na}+$$^{41}\mathrm{K}$}.
	
	Moreover, the observation of the inelastic loss feature deviating significantly from the corresponding elastic peak indicates that careful theoretical investigation is recommended in case unexpected deviations appear in an analysis of a performed Feshbach resonance atom-loss measurement.
	
	The measurements and refined molecular potentials will greatly aid the future investigation of interspin phenomena such as droplet formation \cite{DropletTheoryPetrov2015} as well as in producing stable $^{23}\text{Na}^{39}\mathrm{K}$ molecules in their absolute ground state via a STIRAP process \cite{STIRAPBasicPaper1988}.

	\section*{Acknowledgements}
	
	We gratefully acknowledge financial support from the European Research Council through ERC Starting Grant POLAR and from the Deutsche Forschungsgemeinschaft (DFG) through CRC 1227 (DQ-mat), project A03 and FOR2247, project E5. K.K.V. thanks the Deutsche Forschungsgemeinschaft for financial support through Research Training Group 1991.
	
	\pagebreak

			\newpage\null\thispagestyle{empty}\newpage

			\appendix
				\section{Molecular potential curves}
			
			The parametrization of the molecular potentials is described for example in \cite{Gerdes2008Ap}. The potentials are represented in three parts: the repulsive short-range part $U_{\mathrm{SR}}(R)$, the intermediate range $U_{\mathrm{IR}}(R)$ and the asymptotic long range part $U_{\mathrm{LR}}(R)$, which are given by the following expressions:
			\begin{equation}
			U_{\mathrm{SR}}(R)=A+\frac{B}{R^q}\hspace{0.6cm}\mathrm{for}\hspace{0.6cm} R<R_i ,
			\end{equation}
			\begin{equation}
			U_{\mathrm{IR}}(R)=\sum_{k=0}^n a_k \xi(R)^k \hspace{0.6cm}\mathrm{for}\hspace{0.6cm} R_i\leq R\leq R_o ,
			\end{equation}
			\begin{equation}
			\mathrm{with}\hspace{0.2cm} \xi(R)=\frac{R-R_m}{R+b R_m} , \\
			\end{equation}
			and
			\begin{equation}
			U_{\mathrm{LR}}(R)=-\frac{C_6}{R^6}-\frac{C_8}{R^8}-\frac{C_{10}}{R^{10}}-...\pm E_{\mathrm{ex}}\hspace{0.2cm}\mathrm{for}\hspace{0.2cm} R<R_o, 
			\end{equation}
			where the exchange energy is given by
			\begin{equation}
			E_{\mathrm{ex}}=A_{\mathrm{ex}} R^\gamma \exp{(-\beta R)}.
			\end{equation}
			It is negative for the singlet and positive for the triplet potential.

			The parameters of the refined NaK singlet and triplet molecular potential curves are listed in Tab.~\ref{tab:NaKpot}. A computer code in FORTRAN for calculating the potential functions can be found in the supplement of \cite{Pires2014a}.
			
			\section{Evaluation of Feshbach spectroscopy}
			
			We present a detailed evaluation of 82 independent data points obtained in Feshbach spectroscopy and taken from different publications. The quality of four different PEC-representations is compared in terms of the sum of weighted squared deviations in Tab.~\ref{tab:82Resonances}. Note that the reference \cite{VielSimoniAp} only takes a subset of the given scattering features into account. Therefore, a sum of weighted squared deviations is only calculated for the last three columns of Tab.~\ref{tab:82Resonances}.
			
			\newpage
			\begin{table*}[t]
				\caption{Potential parameters of the $X^1\Sigma^+$ and $a^3\Sigma^+$ states of NaK given with respect to the Na(3s)+K(4s) asymptote.}
				\label{tab:NaKpot}  
				\begin{center}
					\begin{tabular}{lr|lr}
						\hline
						\hline
						\multicolumn{2}{c|}{$X^{1}\Sigma^{+}$} & \multicolumn{2}{c}{$a^{3}\Sigma^{+}$} \\ \hline
						\multicolumn{2}{c}{For $R < R_{i} = 2.617 \textnormal{\AA} $} & \multicolumn{2}{|c}{For $R < R_{i} = 4.60 \textnormal{\AA} $}\\
						
						A & -0.7205185 $\times 10^{4}$ cm$^{-1}$ &A & -0.13272988 $\times 10^{4}$ cm$^{-1}$ \\
						B & 0.696124608 $\times 10^{6}$ cm$^{-1}$ $\textnormal{\AA}^q$ & B &  0.212875317 $\times 10^{5}$ cm$^{-1}$ $\textnormal{\AA}^q$\\
						q & 4.92948 & q & 1.844150\\
						\hline
						\multicolumn{2}{c|}{For $ R_{i} \leq R \leq R_{o} $} & \multicolumn{2}{|c}{For $ R_{i} \leq R \leq R_{o} $}\\
						
						b & -0.4 & b & -0.27\\
						$R_m$ & 3.49901422 \AA & $R_m$ & 5.4478206 \AA\\
						$a_0$    &  -5273.62315 cm$^{-1}$                              & $a_0$    & -207.81119 cm$^{-1}$\\
						$a_1$    &  -0.239542348630413837  $\times 10^{ 1}$  cm$^{-1}$ & $a_1$    & -0.474837907736683607                  cm$^{-1}$\\                        
						$a_2$    &   0.145382657416013644  $\times 10^{ 5}$  cm$^{-1}$ & $a_2$    &  0.178992974113576133 $\times 10^{4}$  cm$^{-1}$\\    
						$a_3$    &   0.114848641509625941  $\times 10^{ 5}$  cm$^{-1}$ & $a_3$    & -0.159602468357546013 $\times 10^{4}$  cm$^{-1}$\\    
						$a_4$    &  -0.393070200439200050  $\times 10^{ 3}$  cm$^{-1}$ & $a_4$    & -0.948541718924311908 $\times 10^{3}$  cm$^{-1}$\\    
						$a_5$    &  -0.169145814548076414  $\times 10^{ 5}$  cm$^{-1}$ & $a_5$    & -0.135179373273532747 $\times 10^{4}$  cm$^{-1}$\\    
						$a_6$    &  -0.374171063602873910  $\times 10^{ 5}$  cm$^{-1}$ & $a_6$    & -0.183565449370861752 $\times 10^{5}$  cm$^{-1}$\\    
						$a_7$    &   0.106844724280541472  $\times 10^{ 6}$  cm$^{-1}$ & $a_7$    &  0.124501710356527314 $\times 10^{6}$  cm$^{-1}$\\    
						$a_8$    &   0.549571543607791886  $\times 10^{ 6}$  cm$^{-1}$ & $a_8$    & -0.163160543217713166 $\times 10^{5}$  cm$^{-1}$\\    
						$a_9$    &  -0.216398544375193026  $\times 10^{ 7}$  cm$^{-1}$ & $a_9$    & -0.199688039882199257 $\times 10^{7}$  cm$^{-1}$\\    
						$a_{10}$ &  -0.101610099703415297  $\times 10^{ 8}$  cm$^{-1}$ & $a_{10}$ &  0.617100814516823366 $\times 10^{7}$  cm$^{-1}$\\    
						$a_{11}$ &   0.221444819359695017  $\times 10^{ 8}$  cm$^{-1}$ & $a_{11}$ &  0.588039077124197735 $\times 10^{6}$  cm$^{-1}$\\    
						$a_{12}$ &   0.109959157819038272  $\times 10^{ 9}$  cm$^{-1}$ & $a_{12}$ & -0.391885588318469822 $\times 10^{8}$  cm$^{-1}$\\    
						$a_{13}$ &  -0.154974082312119037  $\times 10^{ 9}$  cm$^{-1}$ & $a_{13}$ &  0.881312470507461876 $\times 10^{8}$  cm$^{-1}$\\    
						$a_{14}$ &  -0.782460601529465795  $\times 10^{ 9}$  cm$^{-1}$ & $a_{14}$ & -0.839469806952623278 $\times 10^{8}$  cm$^{-1}$\\    
						$a_{15}$ &   0.764737042077244759  $\times 10^{ 9}$  cm$^{-1}$ & $a_{15}$ &  0.307023775214641131 $\times 10^{8}$  cm$^{-1}$\\    
						$a_{16}$ &   0.381868029858328533  $\times 10^{10}$  cm$^{-1}$ & &\\   
						$a_{17}$ &  -0.270560975156805658  $\times 10^{10}$  cm$^{-1}$ & &\\   
						$a_{18}$ &  -0.130777134652790947  $\times 10^{11}$  cm$^{-1}$ & &\\   
						$a_{19}$ &   0.693123967590401554  $\times 10^{10}$  cm$^{-1}$ & &\\   
						$a_{20}$ &   0.317969910129808044  $\times 10^{11}$  cm$^{-1}$ & &\\   
						$a_{21}$ &  -0.127583274381506557  $\times 10^{11}$  cm$^{-1}$ & &\\   
						$a_{22}$ &  -0.547443981078124619  $\times 10^{11}$  cm$^{-1}$ & &\\   
						$a_{23}$ &   0.164038438389521656  $\times 10^{11}$  cm$^{-1}$ & &\\   
						$a_{24}$ &   0.653485806778233261  $\times 10^{11}$  cm$^{-1}$ & &\\   
						$a_{25}$ &  -0.139350456346844196  $\times 10^{11}$  cm$^{-1}$ & &\\   
						$a_{26}$ &  -0.514892853898448334  $\times 10^{11}$  cm$^{-1}$ & &\\   
						$a_{27}$ &   0.700668473929830647  $\times 10^{10}$  cm$^{-1}$ & &\\   
						$a_{28}$ &   0.240948997045685349  $\times 10^{11}$  cm$^{-1}$ & &\\   
						$a_{29}$ &  -0.157575108054349303  $\times 10^{10}$  cm$^{-1}$ & &\\    
						$a_{30}$ &  -0.507254397888037300  $\times 10^{10}$  cm$^{-1}$ & &\\    
						\hline                                                         
						\multicolumn{2}{c|}{For $R > R_{o} = 11.30 \textnormal{\textnormal{\AA}} $} &\multicolumn{2}{|c}{For $R > R_{o} = 11.3 \textnormal{\textnormal{\AA}} $}\\
						
						$C_6$ & 0.1184012$\times 10^{8}$ cm$^{-1}$ $\textnormal{\textnormal{\AA}}^ 6$ & $C_6$ & 0.1184012$\times 10^{8}$ cm$^{-1}$ $\textnormal{\textnormal{\AA}}^ 6$\\
						$C_8$ & 0.3261886$\times 10^{9}$ cm$^{-1}$ $\textnormal{\textnormal{\AA}}^8$ & $C_8$ & 0.3261886$\times 10^{9}$ cm$^{-1}$ $\textnormal{\textnormal{\AA}}^8$\\
						$C_{10}$ & 0.6317249$\times 10^{10}$ cm$^{-1}$ $\textnormal{\textnormal{\AA}}^{10}$ & $C_{10}$ & 0.6317249$\times 10^{10}$ cm$^{-1}$ $\textnormal{\textnormal{\AA}}^{10}$\\
						$A_{ex}$ &0.41447134$\times 10^{4}$ cm$^{-1}$ $\textnormal{\textnormal{\AA}}^{-\gamma}$ & $A_{ex}$ &0.41447134$\times 10^{4}$ cm$^{-1}$ $\textnormal{\textnormal{\AA}}^{-\gamma}$ \\
						$\gamma$ & 5.25669 & $\gamma$ & 5.25669\\
						$\beta$ & 2.11445 $\textnormal{\textnormal{\AA}}^{-1}$ & $\beta$ & 2.11445 $\textnormal{\textnormal{\AA}}^{-1}$\\
						\hline
						\hline
					\end{tabular}
				\end{center}
			\end{table*}

			\newpage
			\begin{table*}[t]
				\caption{Listed are 82 scattering features measured with Feshbach spectroscopy by different groups (see column "ref"). The "note" column contains information about the scattering feature. $ZC$ refers to a zero crossing of the scattering length, $in$ to a loss feature which results from an inelastic two-body collision process, $ov$ indicates the presence of two overlapping structures and an empty entry refers to a Feshbach resonance. $M$ is the total magnetic quantum number of the the entrance channel listed in column $|f,m_f\rangle$. $l$ and $l_{max}$ give the partial waves considered in the calculation of the feature position. $B_{exp}$ gives the experimentally determined position of the scattering feature and $\sigma_{exp}$ the error. The difference $\Delta$ of experimentally determined and calculated positions of the features for four different representations for the PECs are compared in the last four columns of the table. Subscripts indicate the publications the PECs are taken from. The question mark in the $\Delta$(G)\cite{VielSimoniAp} column indicates that the correct assignment is unknown for that specific data point.}
				\label{tab:82Resonances}  
				\begin{center}
					\begin{tabular}{c c c c c c c c c c c | c c c c}

						\multicolumn{2}{c}{Isotope} & \multirow{ 2}{*}{note} & \multirow{ 2}{*}{\,\,$M$\,\,} &\multicolumn{2}{c}{$|f,m_f\rangle$}& \multirow{ 2}{*}{\,\,$l$\,\,} & \multirow{ 2}{*}{$l_{max}$}& \multirow{ 2}{*}{\,\,$B_{exp}(G)$\,\,}& \multirow{ 2}{*}{$\sigma_{exp}$(G)} & \multirow{ 2}{*}{\,\,\,\,ref\,\,\,\,} & \multirow{ 2}{*}{$\Delta$(G)\cite{VielSimoniAp}} & \multirow{ 2}{*}{$\Delta$(G)\cite{STIRAP23Na39KhotbeamstudyTemelkov}} & \multirow{ 2}{*}{$\Delta$(G)\cite{FBR23Na40K2017ChineseGroup}}& \multirow{ 2}{*}{$\Delta$(G)\cite{thispaper}} \\
						\,\,Na\,\, & \,\,K\,\, & & & \,\,\,\,Na\,\,\,\, & K & & & & & & & & &\\ 
						\hline
						\hline
						23&39& & 2.0&(1,1)&(1,1) &0&0&411.334 &1.276& \cite{thispaper}  &-31.18&-1.334&1.206&1.245 \\ 
						23&39& & 2.0&(1,1)&(1,1) &0&0&508.730 &0.831& \cite{thispaper} & -27.27&-2.486&-0.330&-0.079 \\ 
						23&39&\textit{ZC}&2.0&(1,1)&(1,1) &0&0&380.88  &3.83& \cite{thispaper}&-24.14&-2.439&-0.614&-0.545\\ 
						23&39& & 1.0&(1,1)&(1,0) &0&0&6.716 &2.089& \cite{thispaper}  &-28.98&-0.889&1.082&0.103 \\ 
						23&39& & 1.0&(1,1)&(1,0) &0&0&329.115  &0.772& \cite{thispaper}& -27.10&-2.043&-0.019&0.157 \\ 
						23&39& & 1.0&(1,1)&(1,0) &0&0&579.940  &0.880& \cite{thispaper} &-26.58&-2.841&-0.778&-0.550\\ 
						23&39& & 0.0&(1,0)&(1,0) &0&0&5.473 &1.011 & \cite{thispaper}& - &-0.741&0.396&-0.124 \\ 
						23&39& & 0.0&(1,0)&(1,0) &0&0&31.860 &1.685 & \cite{thispaper}&-1.74&-0.850&4.468&2.030\\  
						23&39&\textit{in}&0.0&(1,1)&(1,-1)&0&0&26.340 &3.311& \cite{thispaper} & - &-3.500&-0.820&-1.860 \\ 
						23&39& & 0.0&(1,1)&(1,-1)&0&0&393.608&0.759 & \cite{thispaper} & -28.90&-2.372&-0.182&0.019\\  
						23&39& & 0.0&(1,1)&(1,-1)&0&0&536.066 &0.940& \cite{thispaper} & -29.99&-3.007&-0.702&-0.608 \\ 
						23&39&\textit{ZC}&0.0&(1,1)&(1,-1)&0&0&515.854 &1.684& \cite{thispaper}  &-23.95&-2.490&-0.669&-0.558 \\ 
						23&39& &-1.0&(1,-1)&(1,0)&0&0&245.764 &1.446& \cite{thispaper} & - & 1.046&0.550&1.015 \\ 
						23&39& &-1.0&(1,0)&(1,-1)&0&0&541.087&1.496 & \cite{thispaper} & - &-2.657&0.751&0.564 \\ 
						23&39& &-2.0&(1,0)&(2,-2)&0&0&228.481  &1.492& \cite{thispaper}& -129?&-2.831&-0.285&-1.026\\  
						23&39&\textit{ZC}&-2.0&(1,0)&(2,-2)&0&0&570.286  &2.546 & \cite{thispaper}&-29.15&-5.947&-4.033&-4.021 \\ 
						23&39& &-2.0&(1,-1)&(1,-1)&0&0&32.475 &0.830& \cite{SchulzeBEC2018Ap} &30.47&1.043&-1.643&-0.663 \\ 
						23&39& &-2.0&(1,-1)&(1,-1)&0&0&247.108&0.230& \cite{SchulzeBEC2018Ap} &5.71&-0.329&-0.969&-0.460 \\ 
						23&39&\textit{ZC}&-2.0&(1,-1)&(1,-1)&0&0&117.189&0.150& \cite{SchulzeBEC2018Ap} &41.48&3.405&0.024&0.109\\  
						23&39&\textit{ov} &-2.0&(1,-1)&(1,-1)&0&0&646.600&1.500& \cite{SchulzeBEC2018Ap} & - & -&-3.800&-4.585\\ 
						\hline
						23&40& &-3.5&(1,1)&(9/2,-9/2)&0&0&78.320 &0.150& \cite{FBR23Na40K2012} &0.54&0.008&-0.066&-0.024 \\ 
						23&40& &-3.5&(1,1)&(9/2,-9/2)&0&0&89.700 & 0.250&\cite{FBR23Na40K2012,RuiStuff}&1.02&0.692&0.596&-0.084 \\ 
						23&40& &-2.5&(1,1)&(9/2,-7/2)&0&0&81.620 &0.160 &\cite{FBR23Na40K2012} &0.20&-0.108&-0.160&-0.032 \\ 
						23&40& &-2.5&(1,1)&(9/2,-7/2)&0&0&89.780 &0.460 &\cite{FBR23Na40K2012} &-0.04&-0.594&-0.672&-0.609 \\ 
						23&40& &-2.5&(1,1)&(9/2,-7/2)&0&0&108.600 &3.000 &\cite{FBR23Na40K2012} &-0.31&-0.606&-0.719&-1.694 \\ 
						23&40& &-1.5&(1,1)&(9/2,-5/2)&0&0&96.540 &0.090 &\cite{FBR23Na40K2012}&0.15&-0.091&-0.141&0.002 \\ 
						23&40& &-1.5&(1,1)&(9/2,-5/2)&0&0&106.920 &0.270& \cite{FBR23Na40K2012} &0.38&-0.155&-0.234&-0.152 \\ 
						23&40& &-1.5&(1,1)&(9/2,-5/2)&0&0&138.560 & 1.000&\cite{FBR23Na40K2012,RuiStuff}&1.74&1.518&1.384&0.042 \\ 
						23&40& &-0.5&(1,1)&(9/2,-3/2)&0&0&116.910 &0.150& \cite{FBR23Na40K2012} &-0.28&-0.350&-0.388&-0.226 \\ 
						23&40& &-0.5&(1,1)&(9/2,-3/2)&0&0&130.640 & 0.030&\cite{FBR23Na40K2012,RuiStuff}&0.28&-0.127&-0.199&-0.100 \\ 
						23&40& &-0.5&(1,1)&(9/2,-3/2)&0&0&175.000  &5.000& \cite{FBR23Na40K2012}&-2.44&-2.520&-2.674&-4.531 \\ 
						23&40& &-4.5&(1,1)&(9/2,-9/2)&1&1&6.350 &0.030& \cite{FBR23Na40K2012} &-0.15&-0.034&-0.042&-0.030 \\ 
						23&40& &-2.5&(1,1)&(9/2,-9/2)&1&1&6.410 &0.030& \cite{FBR23Na40K2012} &-0.16&0.003&-0.008&-0.002 \\ 
						23&40& &-3.5&(1,1)&(9/2,-9/2)&1&1&6.470 &0.030& \cite{FBR23Na40K2012} &-0.16&0.006&-0.004&-0.002 \\ 
						23&40& &-3.5&(1,1)&(9/2,-9/2)&1&1&6.680 &0.030& \cite{FBR23Na40K2012} &-0.17&0.013&0.005&0.010 \\ 
						23&40& &-2.5&(1,1)&(9/2,-9/2)&1&1&19.100 &0.100& \cite{FBR23Na40K2012} &-0.22&-0.056&-0.060&-0.052 \\ 
						23&40& &-4.5&(1,1)&(9/2,-9/2)&1&1&19.200 &0.100& \cite{FBR23Na40K2012} &-0.19&0.016&0.012&0.008 \\ 
						23&40& &-3.5&(1,1)&(9/2,-9/2)&1&1&19.300 &0.100& \cite{FBR23Na40K2012} &-0.17&0.048&0.044&0.036 \\ 
						23&40& &-3.5&(1,1)&(9/2,-7/2)&1&1&7.320 &0.140& \cite{FBR23Na40K2012} &-0.30&-0.119&-0.133&-0.134 \\ 
						23&40& &-1.5&(1,1)&(9/2,-7/2)&1&1&7.540 &0.040& \cite{FBR23Na40K2012} &-0.36&-0.148&-0.160&-0.156 \\ 
						23&40& &-2.5&(1,1)&(9/2,-7/2)&1&1&7.540 &0.040& \cite{FBR23Na40K2012} &-0.36&-0.164&-0.176&-0.168 \\ 
						23&40& &-3.5&(1,1)&(9/2,-7/2)&1&1&7.540 &0.040& \cite{FBR23Na40K2012} &-0.34&-0.138&-0.150&-0.144 \\ 
						23&40& &-2.5&(1,1)&(9/2,-7/2)&1&1&23.190 &0.040& \cite{FBR23Na40K2012} &-0.30&-0.048&-0.048&-0.048 \\ 
						23&40& &-3.5&(1,1)&(9/2,-7/2)&1&1&23.290 &0.040& \cite{FBR23Na40K2012} &-0.31&-0.012&-0.012&-0.024 \\ 
						23&40&\textit{in}&-2.5&(1,1)&(9/2,-5/2)&1&1&9.230  &0.340& \cite{FBR23Na40K2012}&-0.11&0.108&0.092&0.096 \\ 
						23&40& &-1.5&(1,1)&(9/2,-5/2)&1&1&9.600 &0.040& \cite{FBR23Na40K2012} &-0.29&0.032&0.016&0.028 \\ 
						23&40& &-2.5&(1,1)&(9/2,-5/2)&1&1&9.600 &0.040& \cite{FBR23Na40K2012} &-0.28&0.004&-0.012&-0.004 \\ 
						23&40& &-0.5&(1,1)&(9/2,-5/2)&1&1&9.600 &0.040& \cite{FBR23Na40K2012} &-0.29&-0.004&-0.016&-0.012 \\ 
						23&40& &-1.5&(1,1)&(9/2,-5/2)&1&1&29.200 &0.090& \cite{FBR23Na40K2012} &-0.41&-0.080&-0.072&-0.064 \\

					\end{tabular}
					
				\end{center}
			\end{table*}
			
			\newpage
			
			\begin{table*}[t]
				
				\begin{center}
					\begin{tabular}{c c c c c c c c c c c| c c c c}
						\multicolumn{2}{c}{Isotope} & \multirow{ 2}{*}{note} & \multirow{ 2}{*}{\,\,$M$\,\,} &\multicolumn{2}{c}{$|f,m_f\rangle$}& \multirow{ 2}{*}{\,\,$l$\,\,} & \multirow{ 2}{*}{$l_{max}$}& \multirow{ 2}{*}{\,\,$B_{exp}$(G)\,\,}& \multirow{ 2}{*}{$\sigma_{exp}$(G)} & \multirow{ 2}{*}{\,\,\,\,ref\,\,\,\,} & \multirow{ 2}{*}{$\Delta$(G)\cite{VielSimoniAp}} & \multirow{ 2}{*}{$\Delta$(G)\cite{STIRAP23Na39KhotbeamstudyTemelkov}} & \multirow{ 2}{*}{$\Delta$(G)\cite{FBR23Na40K2017ChineseGroup}}& \multirow{ 2}{*}{$\Delta$(G)\cite{thispaper}} \\
						\,\,Na\,\, & \,\,K\,\, & & & \,\,\,\,Na\,\,\,\, & K & & & & & & & & &\\ 
						\hline
						\hline
						23&40& &-2.5&(1,1)&(9/2,-5/2)&1&1&29.520 &0.090& \cite{FBR23Na40K2012} &-0.41&0.041&0.046&0.024 \\ 
						23&40& &-0.5&(1,1)&(9/2,-5/2)&1&1&29.450 &0.090& \cite{FBR23Na40K2012} &-0.43&-0.002&0.004&-0.008 \\ 
						23&40& &-0.5&(1,1)&(9/2,-3/2)&1&1&12.510 &0.050& \cite{FBR23Na40K2012} &-0.53&-0.171&-0.183&-0.159 \\ 
						23&40& & 0.5&(1,1)&(9/2,-3/2)&1&1&12.680 &0.050& \cite{FBR23Na40K2012} &-0.52&-0.096&-0.113&-0.108 \\ 
						23&40& &-0.5&(1,1)&(9/2,-3/2)&1&1&39.390 &0.040& \cite{FBR23Na40K2012} &-0.48&0.059&0.080&0.100 \\ 
						23&40& & 0.5&(1,1)&(9/2,-3/2)&1&1&39.850 &0.040& \cite{FBR23Na40K2012} &-0.52&0.192&0.215&0.188 \\ 
						23&40& & 0.5&(1,1)&(9/2,-1/2)&0&0&146.700 &0.200& \cite{FBR23Na40K2017ChineseGroup} & - &-0.251&-0.245&-0.065 \\ 
						23&40& & 0.5&(1,1)&(9/2,-1/2)&0&0&165.340 &0.300& \cite{FBR23Na40K2017ChineseGroup} & - &-0.340&-0.383&-0.269 \\ 
						23&40& & 0.5&(1,1)&(9/2,-1/2)&0&0&233.000 &1.800& \cite{FBR23Na40K2017ChineseGroup} & - &-4.994&-5.126&-7.722 \\ 
						23&40& & 1.5&(1,1)&(9/2, 1/2)&0&0&190.500 &0.200& \cite{FBR23Na40K2017ChineseGroup} & - &-0.620&-0.513&-0.319 \\ 
						23&40& & 1.5&(1,1)&(9/2, 1/2)&0&0&218.400 &0.200& \cite{FBR23Na40K2017ChineseGroup} & - &-0.667&-0.622&-0.505 \\ 
						23&40& & 1.5&(1,1)&(9/2, 1/2)&0&0&308.100 &3.220& \cite{FBR23Na40K2017ChineseGroup} & - &-19.413 & -19.588&-23.114\\  
						23&40& & 2.5&(1,1)&(9/2, 3/2)&0&0&256.600 &0.200& \cite{FBR23Na40K2017ChineseGroup} & - &-0.965&-0.619&-0.423 \\ 
						23&40& & 2.5&(1,1)&(9/2, 3/2)&0&0&299.900 &0.400& \cite{FBR23Na40K2017ChineseGroup} & - &-1.900&-1.630&-1.522 \\ 
						23&40& & 0.5&(1,1)&(9/2,-1/2)&1&1&18.810 &0.100& \cite{FBR23Na40K2017ChineseGroup} & - &-0.024&-0.031&0.012 \\ 
						23&40& & 1.5&(1,1)&(9/2,-1/2)&1&1&19.150 &0.100& \cite{FBR23Na40K2017ChineseGroup} & - &0.112&0.107&0.116 \\ 
						23&40& & 0.5&(1,1)&(9/2,-1/2)&1&1&58.320 &0.100& \cite{FBR23Na40K2017ChineseGroup} & - &-0.103&-0.036&-0.012\\  
						23&40& & 1.5&(1,1)&(9/2,-1/2)&1&1&59.100 &0.100& \cite{FBR23Na40K2017ChineseGroup} & - &0.184&0.249&0.200 \\ 
						23&40& & 1.5&(1,1)&(9/2, 1/2)&1&1&35.170 &0.100& \cite{FBR23Na40K2017ChineseGroup} & - &-0.156&-0.116&-0.025\\  
						23&40& & 2.5&(1,1)&(9/2, 1/2)&1&1&35.830 &0.100& \cite{FBR23Na40K2017ChineseGroup} & - &0.047&0.084&0.112 \\ 
						23&40& & 1.5&(1,1)&(9/2, 1/2)&1&1&100.360 &0.100& \cite{FBR23Na40K2017ChineseGroup} & - &-0.104&0.111&0.112 \\ 
						23&40& & 2.5&(1,1)&(9/2, 1/2)&1&1&101.310 &0.100& \cite{FBR23Na40K2017ChineseGroup} & - &0.171&0.381&0.285 \\ 
						23&40& &-3.5&(1,1)&(9/2,-9/2)&0&2&204.520 &0.200& \cite{FBR23Na40K2017ChineseGroup} & - &-3.75&0.084&0.116 \\ 
						23&40& &-3.5&(1,1)&(9/2,-9/2)&0&2&279.800 &0.200& \cite{FBR23Na40K2017ChineseGroup} & - &-5.99&-0.202&0.024 \\ 
						23&40&\textit{in}&-2.5&(1,1)&(9/2,-7/2)&0&2&202.680 &0.200& \cite{FBR23Na40K2017ChineseGroup} & - &-3.51&0.105&0.112 \\ 
						23&40&\textit{in}&-2.5&(1,1)&(9/2,-7/2)&0&2&276.300 &0.200& \cite{FBR23Na40K2017ChineseGroup} & - &-5.50&-0.040&0.113 \\ 
						23&40&\textit{in}&-1.5&(1,1)&(9/2,-5/2)&0&2&201.660 &0.200& \cite{FBR23Na40K2017ChineseGroup} & - &-3.29&0.117&0.096 \\ 
						23&40&\textit{in}&-1.5&(1,1)&(9/2,-5/2)&0&2&274.600 &0.200& \cite{FBR23Na40K2017ChineseGroup} & - &-5.14&0.026&0.115 \\ 
						23&40&\textit{in}&-0.5&(1,1)&(9/2,-3/2)&0&2&201.440 &0.200& \cite{FBR23Na40K2017ChineseGroup} & - &-3.06&0.124&0.076 \\ 
						23&40&\textit{in}&-0.5&(1,1)&(9/2,-3/2)&0&2&274.800 &0.200& \cite{FBR23Na40K2017ChineseGroup} & - &-4.67&0.223&0.251 \\ 
						23&40&\textit{in}&0.5&(1,1)&(9/2,-1/2)&0&2&202.100 &0.200& \cite{FBR23Na40K2017ChineseGroup} & - &-2.77&0.168&0.096 \\ 
						23&40&\textit{in}&0.5&(1,1)&(9/2,-1/2)&0&2&276.200 &1.300& \cite{FBR23Na40K2017ChineseGroup} & - &-4.67&-0.048&-0.079 \\ 
						23&40&\textit{in}&1.5&(1,1)&(9/2, 1/2)&0&2&278.800 &0.400& \cite{FBR23Na40K2017ChineseGroup} & - &-5.02&-0.711&-0.795 \\ 
						23&40&\textit{in}&2.5&(1,1)&(9/2, 3/2)&0&2&283.700 &0.900& \cite{FBR23Na40K2017ChineseGroup} & - &-3.98&0.003&-0.160 \\ 
						
						\hline
						\multicolumn{11}{c|}{\textbf{sum of weighted squared deviations:}}& \textbf{-} &\textbf{5240.5}&\textbf{337.0}&\textbf{255.31}\\
						\hline
						\hline
					\end{tabular}
				\end{center}
			\end{table*}


\begin{thebibliography}{32}
\bibitem[{Amp()}]{Amplitude}
		\bibinfo{note}{This procedure is most efficient for a low $^{39}\mathrm{K}$ intraspecies scattering length because self-evaporation is suppressed in this case.}
\end{thebibliography}
		\end{document}